\documentclass{emulateapj}
\usepackage{graphicx}
\usepackage{amssymb}
\usepackage{natbib}
\newcommand{\asec}      {\mbox{$^{\prime \prime}  $} }

\begin{document}
\title{Spectroscopic Confirmation Of An Extreme Starburst At Redshift 4.547{*}}
\author{Peter Capak\altaffilmark{1,2}, 
C.L. Carilli\altaffilmark{3},
N. Lee \altaffilmark{3},
T. Aldcroft\altaffilmark{4},
H. Aussel\altaffilmark{5},
E. Schinnerer\altaffilmark{6}, 
G.W. Wilson\altaffilmark{7},
M.S. Yun\altaffilmark{7}, 
A. Blain\altaffilmark{2},
M. Giavalisco\altaffilmark{7}, 
O. Ilbert\altaffilmark{8}, 
J. Kartaltepe\altaffilmark{8}, 
K.-S. Lee\altaffilmark{9}, 
H. McCracken\altaffilmark{10}, 
B. Mobasher \altaffilmark{11}, 
M. Salvato\altaffilmark{2},
S. Sasaki\altaffilmark{12},
K.S. Scott\altaffilmark{7},
K. Sheth\altaffilmark{1,2},
Y. Shioya\altaffilmark{12},
D. Thompson\altaffilmark{13},
M. Elvis\altaffilmark{4},
D.B. Sanders\altaffilmark{8},  
N.Z. Scoville\altaffilmark{2}
Y.Tanaguchi\altaffilmark{12}}

\altaffiltext{1}{Spitzer Science Center, 314-6 Caltech, Pasadena, CA, 91125}
\altaffiltext{2}{105-24 Caltech, Pasadena, CA, 91125}
\altaffiltext{3}{National Radio Astronomy Observatory, P.O. Box O, Socorro, NM, 87801}
\altaffiltext{4}{Harvard-Smithsonian Center for Astrophysics, 60 Garden Street, Cambridge, MA, 02138}
\altaffiltext{5}{AIM Ð UnitŽ Mixte de Recherche CEA Ð CNRS Ð UniversitŽ Paris VII Ð UMR n¡ 7158}
\altaffiltext{6}{Max Planck Institut f\"{u}r Astronomie, K\"{o}nigstuhl 17, Heidelberg, D-69117, Germany}
\altaffiltext{7}{Astronomy Department, University of Massachusetts, Amherst, MA, 01003, USA}
\altaffiltext{8}{Institute for Astronomy, University of Hawaii, 2680 Woodlawn Drive, Honolulu, HI, 96822, USA}
\altaffiltext{9}{Department of Astronomy, 260 Whitney Ave, Yale University, New Haven, CT, 06511, USA}
\altaffiltext{10}{Institut d'Astrophysique de Paris, UMR7095 CNRS, Universit\`{e} Pierre et Marie Curie, 98 bis Boulevard Arago, 75014 Paris, France}
\altaffiltext{11}{Department of Physics and Astronomy, University of California, Riverside, CA, 92521, USA}
\altaffiltext{12}{Research Center for Space and Cosmic Evolution, Ehime University, 2-5 Bunkyo-cho, Matsuyama 790-8577, Japan}
\altaffiltext{13}{Large Binocular Telescope Observatory, University of Arizona, 933 N. Cherry Ave., Tucson, AZ, 85721, USA}
\altaffiltext{*}{Based on observations taken at the Keck Observatory, the James Clerk Maxwell Telescope, the Institut de Radioastronomie Millimetrique 30m telescope, the Galaxy Evolution Explorer, the Chandra X-ray Observatory, the Hubble Space Telescope, the Very Large Array, the Subaru Telescope, the United Kingdom Infrared Telescope, and the Canada France Hawaii Telescope.}


\begin{abstract}
We report the spectroscopic confirmation of a sub-mm galaxy (SMG) at $z=4.547$ with an estimated L$_{IR}=0.5-2.0\times10^{13}L_\sun$.  The spectra, mid-IR, and X-ray properties  indicate the bolometric luminosity is  dominated by star formation at a rate of $>1000$M$_\sun yr^{-1}$.  Multiple, spatially separated components are visible in the Ly-Alpha line with an observed velocity difference of up to $380$ km/sec and the object morphology indicates a merger.   The best fit spectral energy distribution and spectral line indicators suggest the object is 2-8 Myr old and contains $>10^{10}$M$_\sun$ of stellar mass.  This object is a likely progenitor for the massive early type systems seen at $z\sim2$.
\end{abstract}

\section{Introduction \label{s:intro}}
The study of galaxies detected at millimeter (mm) and sub-mm wavelengths is one of the most rapidly developing fields in observational astronomy.  It is now known that a large fraction of the star formation activity is enshrouded in dust, with the star formation rate (SFR) being directly proportional to the far infrared (FIR) luminosity of galaxies, modulo possible contributions from an Active Galactic Nuclei (AGN)\citep{1998Natur.394..241H}.   Surveys performed at mm wavelengths directly probe the FIR luminosity, and hence the amount of star formation.  Furthermore, the shape of the galaxy spectral energy distributions (SEDs) at rest-frame mm wavelengths results in a negative K-correction between $0.5<z<10$.  Therefore a flux limited survey is equivalent to a SFR limited survey at these redshifts \citep{2002PhR...369..111B}. 

The current redshift distribution of mm galaxies peaks at $z\sim 2$, with very few galaxies at $z > 3$  \citep{2005ApJ...622..772C, 2005MNRAS.358..149P, 2007MNRAS.379.1571A}.  However, the small bandwidth of current mm wave spectrographs makes it very difficult to measure redshifts directly, and the low angular resolution of mm single dish imaging leads to multiple optical candidates for the same source.  As a result, mm surveys have relied on high resolution radio data to identify the optical counterparts for subsequent spectroscopic follow-up. This leaves  $35-70$\% of the population of mm-selected galaxies at milli-Jansky flux levels unidentified and potentially at higher redshift \citep{2007ApJ...670L..89W, 2007ApJ...671.1531Y}. 

\begin{figure*}
\begin{center}
\includegraphics{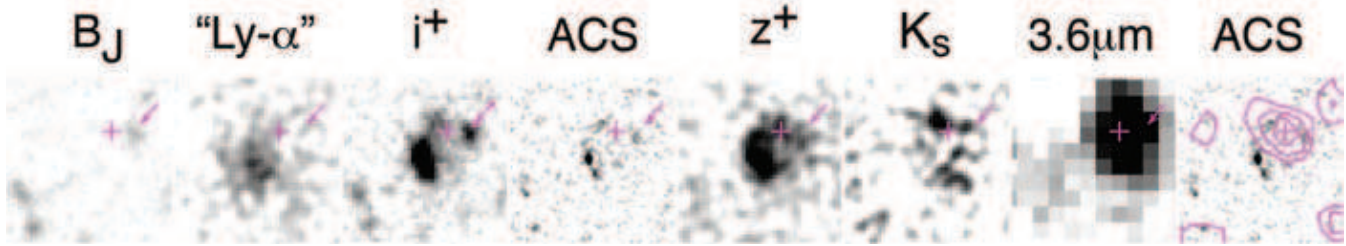}
\caption{Images in $B_J$, Ly-$\alpha$, $i^+$, ACS F814W, $z^+$, $K_s$, and 3.6$\mu$m are shown for $6\asec \times 6\asec$ (39.4kpc) around the source.  The radio position is marked with a cross and a foreground object is marked with an arrow.  1.4GHz Radio contours are overlaid on the $ACS$ band image in the right most panel.   The Ly-$\alpha$ image is generated by subtracting the $r^+$ continuum image from the IA679 image which is centered on the Ly$-\alpha$ line.  The 3.6$\mu$m (rest frame optical) flux is centered on the radio position and diffuse $i^+$ and $z^+$ (rest frame UV) flux.  A UV bright knot is visible down and to the left (0.6\asec to the SE) of the radio position and contains (73\%) of the UV flux.  The Ly-$\alpha$ emission is more extended than the UV emission seen in the $i^+$ band and originates from both the UV bright knot and the radio position but not the foreground objects.  Note the UV bright knot is absent in the $K_s$ band image, indicating a very young age. 
\label{f:ground}}
\end{center}
\end{figure*}

\begin{figure}
\begin{center}
\includegraphics[scale=1]{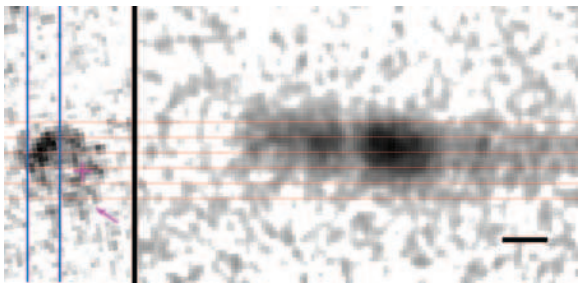}
\caption{The 2-D spectra of the object around the Ly-$\alpha$ line is shown on the right along with the  ground based $z^{+}$ band image on the left with the slit position marked in blue, the radio position marked with a magenta cross, and the foreground object indicated by a magenta arrow.  The red lines are spaced at 0.4\asec intervals which is comparable to the spacial resolution of the spectra and the black bar in the bottom right has a length of 1\AA\ in the rest frame.   Ly-$\alpha$ emission is coming from all portions of the source and there is a clear velocity gradient along the slit.  Note the Ly-$\alpha$ absorption feature at the physical position of the bright continuum emission and the multiple peaks in the Ly-$\alpha$ line.\label{f:2dspec}}
\end{center}
\end{figure}

In this paper we report the discovery of a mm galaxy with a spectroscopic redshift of $z=4.547$ that appears to be dominated by star formation. This is the highest redshift galaxy detected at mm wavelengths not associated with an optically bright quasar.  The object was independently selected as a Lyman-break galaxy \citep{soo-lbg} for spectroscopic followup, a mm source, and a radio source \citep{carilli-stack}.  The source reported here is unusual for mm sources because it has several nearby optically bright counterparts which were selected as V band dropouts and is unusually luminous which allows for a radio detection.  However, the confirmation of this object suggests the population of sources with similar radio to mm flux ratios and optical colors may also be at high redshifts. 

We assume a H$_o=70$, $\Omega_v=0.7$, $\Omega_m=0.3$ cosmology and a star formation rate integrated across a \citet{1955ApJ...121..161S} IMF from 0.1-100$M_\sun$ throughout this paper.

\section{Data}

	Observations at $\lambda=1.1$mm with an average RMS noise of 1.3mJy were obtained with the AzTEC \citep{2008arXiv0801.2783W} camera at the the James Clerk Maxwell Telescope (JCMT) \citep{2008arXiv0801.2779S}.  Additional observations were obtained by the MAMBO camera on the IRAM 30m telescope with an RMS of 0.67 mJy and a positional accuracy of $<5\asec$ (Schinnerer et al. in prep).  Ground based optical and near infrared imaging in 22 bands, Hubble Space Telescope, Spitzer, and Very Large Array images were obtained as part of the COSMOS survey as described in \citet{capak-data}, \citet{scoville-hst}, \citet{sanders-spitzer}, and \citet{eva-vla} respectively.  Additional deep J and $K_s$ data were obtained with the UKIRT, and CFHT telescopes as part of the COSMOS survey (McCraken et al. in prep).  X-ray data were obtained with the Chandra X-ray observatory as part of the C-COSMOS program (Elvis et al. in prep).  The respective fluxes are tabulated in Table \ref{t:fluxes}.

\begin{deluxetable}{lc}
\tabletypesize{\scriptsize}
\tablecaption{Object Flux\label{t:fluxes}}
\tablehead{
\colhead{Wavelength} & \colhead{Flux ($\mu$Jy)\tablenotemark{+}}
}
\startdata
0.2-8Kev	& $<3\times 10^{-16}$ ergs $s^{-1}$ cm$^{-2}$\\
150nm (FUV) & $<0.2$\\
250nm (NUV)& $<0.09$\\
380nm ($u^*$)& $<0.01$\\
427nm & $<0.02$\\
446nm ($B_J$)& $<0.01$\\
464nm & $0.04\pm0.04$\\
478nm ($g^+$)& $0.03\pm0.02$\\
484nm & $0.04\pm0.04$\\
505nm & $0.09\pm0.05$\\
527nm & $0.08\pm0.04$\\
548nm ($V_J$)& $0.15\pm0.02$\\
574nm & $0.18\pm0.05$\\
624nm & $0.25\pm0.06$\\
630nm ($r^+$)& $0.50\pm0.03$\\
679nm & $1.24\pm0.06$\\
709nm & $1.30\pm0.07$\\
711nm & $1.32\pm0.13$\\
738nm & $1.59\pm0.09$\\
764nm ($i^+$) & $1.60\pm0.05$\\
767nm & $1.59\pm0.09$\\
815nm & $1.60\pm0.09$\\
827nm & $1.48\pm0.09$\\
904nm ($z^+$) & $1.82\pm0.10$\\
1.25$\mu$m (J)	&	$2.4\pm0.8$\\
2.15$\mu$m ($K_s$)	&	$3.7\pm0.5$\\
3.6$\mu$m	&	$7.9\pm0.2$\\
4.5$\mu$m	&	$5.8\pm0.4$\\
5.8$\mu$m	&	$3.4\pm1.3$\\
8.0$\mu$m	&	$10\pm3.6$\\
24$\mu$m	&	$26\pm13$\tablenotemark{*}\\
1.1mm	&	$4800\pm1500$\\
1.25mm	&	$3410\pm670$\\
20cm	&	$45\pm9$\\
\enddata
\tablenotetext{+}{All limits are $1\sigma$ limits.}
\tablenotetext{*}{A nearby bright source was modeled and subtracted to make this measurement, the formal upper limit is $70\mu$Jy if the nearby source is not subtracted.}
\end{deluxetable}

\begin{figure}
\begin{center}
\includegraphics[scale=1]{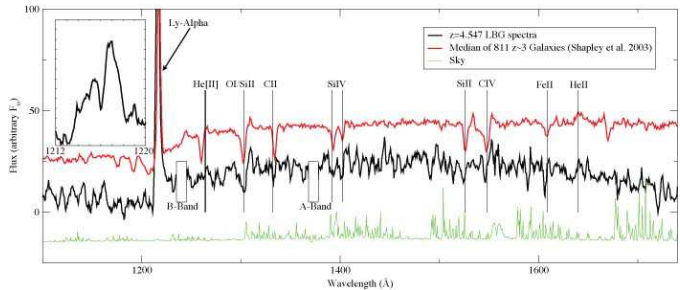}
\caption{The optical spectrum of the source (black) is shown along with the composite LBG spectrum (red) from \citet{2003ApJ...588...65S}.  The subtracted sky spectra is shown in green for comparison and the atmospheric A and B absorption bands are marked with boxes.  The inset panel on the right shows the region around the Ly-$\alpha$ line.  Lines from the interstellar medium and stellar photospheres such as O I,O II,C II,C IV,Si II,Fe II, He II, are clearly visible in the spectra.  The presence of the Si IV 1297\AA\ and C IV 1549\AA\ P-cgyni lines,  along with the He II 1640\AA\ emission lines indicates the presences of both Wolf-Rayet and O stars, placing the age of the burst at a few Myr \citep{2000ApJ...528...96P}.  Note the absence of a common forbidden He[II] absorption feature at 1264\AA\, which is generated in shocked gas, suggesting a we are not looking through a significant column of gas.  \label{f:spectra}}
\end{center}
\end{figure}

	The optical spectra were taken on the Keck-II telescope using the DEIMOS instrument with a 1\asec slit width, the 830 line/mm grating (3.3\AA\ FWHM resolution), and the OG550 blocking filter to optimize red throughput.  The data were collected in eight 1800s exposures (4h total integration) under photometric conditions with 0.4-0.6\asec seeing.  The object was dithered along the slit by $\pm3\asec$ between exposures to improve background subtraction.  The data were reduced with a modified version of the DEEP2 DEIMOS pipeline \citep{2001astro.ph..9164M}.  In addition to the standard processing this modified pipeline constructs and subtracts a median background and accounts for the shifts when combining the spectra.   These additional steps remove the "ghosting" inherent to the 830l/mm grating.  
		 
	 An image of the continuum and Ly-$\alpha$ emission is shown in Figure \ref{f:ground}, and indicates emission from all components of the source.  The 2-D and 1-D spectra are shown in Figures \ref{f:2dspec} and \ref{f:spectra} respectively.  Ly$-\alpha$ emission is detected from both the compact and extended portions of the object with a velocity gradient across the slit.  Ly-$\alpha$ emission from the diffuse region is redshifted with respect to the compact source, and a deep Ly$-\alpha$ absorption feature is present at a redshift corresponding to other absorption features seen in the spectra. The dispersion of distinct peaks in the Ly-$\alpha$ emission is $380$km s$^{-1}$ and the line asymmetry indicates outflow winds of up to $\sim1800$km s$^{-1}$, typical of a merger with heavy star formation.  Several interstellar absorption features are clearly seen in the continuum yielding a best fit redshift of $4.547\pm0.002$,  but the dispersion in the Ly-$\alpha$ line suggests some components could be $\pm0.02$ from the continuum redshift.

	A foreground object with an $i^+$ band flux of $0.58\mu$Jy is visible to the west of our source in the B band image (see Figure \ref{f:ground}) and its spectrum shows [O II] emission at $z=1.41$.  The mass of this system is too low to significantly gravitationally lens our system and it is unlikely that the mm flux originates from this source because the object is outside the radio and Spitzer positional error and has an SED which indicates little obscuration.  Furthermore, if the source of the mm emission is at $z=1.41$ the 24$\mu$m, mm, and radio flux place the far infrared luminosity at $\sim10^{14} L_\sun$ and the dust temperature at less than $20$K, which is physically unreasonable \citep{2005ARA&A..43..677S, 2002ApJ...576..159D, 1999ApJ...513L..13C}. 
	
\section{Pan-Chromatic Properties and Morphology}
	The rest-frame ultraviolet (UV) properties indicate a merger, typical of star-forming mm sources seen at $z\sim2$ \citep{2005ApJ...622..772C, 2005MNRAS.358..149P, 2007MNRAS.379.1571A}.  At least two distinct components are visible in the ACS image, and a region of extended emission is visible in the ground based $i^+$ and $z^+$ images which are more sensitive to extended emission than the HST data.   The rest frame UV is centered at 10:00:54.516, +2:34:35.17 with the radio and rest frame optical emission centered at 10:00:54.48, +2:34:35.9.

\begin{figure}
\begin{center}
\includegraphics[scale=0.35]{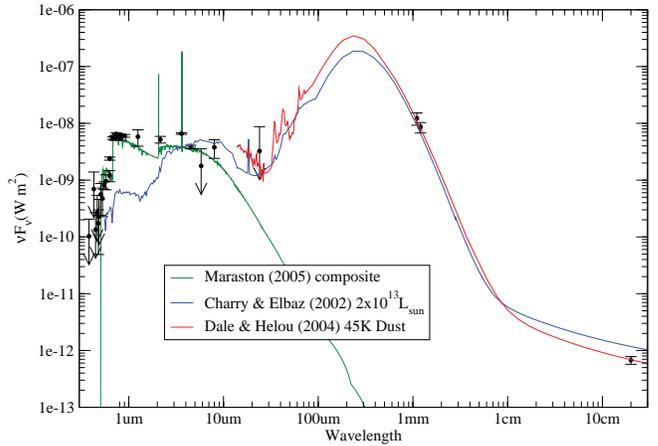}
\caption{The pan-chromatic SED of the source is shown along with the best fit SED model.  The UV bright portion of the source is fit by a 2.5Myr old \citet{2005MNRAS.362..799M} stellar population with $1.9\times10^8$ M$_\sun$ of stellar mass and no extinction, while the UV diffuse portion is fit by a 6.5Myr old population with $9.7\times10^9$ M$_\sun$ of stellar mass and 1.4 magnitudes of visual extinction.  The 8$\mu$m, 24$\mu$m, mm, and radio portion of the SED is fit by a $2\times 10^{13}$L$_\sun$ \citet{2001ApJ...556..562C} model (blue) or a 45K \citet{2002ApJ...576..159D} model (red).  \label{f:sed}}
\end{center}
\end{figure}

\begin{deluxetable}{lcccccl}
\vspace{-0.4in}
\tabletypesize{\scriptsize}
\tablecaption{Best Fit SEDs\label{t:fits}}
\tablehead{
\colhead{Emission} & \colhead{Age} & \colhead{Av} & \colhead{Mass} & \colhead{UV SFR} & \colhead{$\chi^2$} &\colhead{Model}\\
\colhead{Source} & \colhead{(Myr)} & \colhead{} & \colhead{($M_\sun$)} & \colhead{($M_\sun$)} & \colhead{} &\colhead{}
}
\startdata
Diffuse	&	6.5	&	1.4	&	$1\times10^{10}$	&	  $3200\pm550$	&	24.3	& M05\\
		&	7.6	&	1.5	&	$2\times10^{10}$	&	  $4200\pm730$	&	24.7	& BC03\\
Compact	&	2.5	&	0.0	&	$2\times10^8$		&	  $250\pm20$		&	54.5 & M05\\
		&	2.9	&	0.0	&	$6\times10^8$		&	  $250\pm20$		&	55.7 & BC03\\
\enddata
\end{deluxetable}

	The majority (73\%) of the rest frame UV flux originates from a compact ($<2.7$kpc) knot of emission at the south east corner of the object,  with the remaining UV emission extended over 17.7 kpc (2.7\asec), while the rest-frame optical (Spitzer-IRAC) flux of the source is centered on the diffuse UV and radio emission. The IA679 filter corresponds to the rest frame Ly-$\alpha$ line and shows strong emission from all components of the source and an extended Ly-$\alpha$ halo around the source  (see Figure \ref{f:ground}).  In addition, a significant excess of flux is measured in the 3.6$\mu$m band which is centered on the H$\alpha$ line at $z=4.547$.  The Ly-$\alpha$ image, the 3.6$\mu$m excess, and the optical-IR colors of the diffuse region are consistent with all components residing at $z=4.547$.

	 The possible presence of H$\alpha$ in the 3.6$\mu$m band and [O II] in the K$_s$ band combined with the low S/N of the J band and multiple components with different SEDs make it difficult to constrain the age and mass of this object with stellar models because the 4000\AA\ break strength is degenerate with the line ratio in some cases. To reduce these degeneracies H$\alpha$ and [O II] line flux is added to the stellar models in proportion to the unobscured UV star formation rate \citep{1998ARA&A..36..189K}.  The diffuse portion of the source fades rapidly between the K$_s$ and $z^+$ bands, indicating significant obscuration. The UV compact region is not detected redwards of the $z^+$ band, indicating a very young ($<5$ Myr old) stellar population and little dust obscuration.  
	 
	 A two component fit to the total integrated light produced poor results.  The UV and optical light are spatially separated, but the best fit model places the majority of both the UV and optical light in a single un-obscured $>0.1$Gyr old population.  To overcome this degeneracy we attempt to deconvolve the UV bright knot and the diffuse component of the source.  Flux measurements were made in all of the ground based images for the UV diffuse portion of the source using a 1\asec diameter aperture on the original stacked images without PSF matching \citep{capak-data}.  An aperture correction of 1.4 magnitudes, based on the ACS morphology, was then applied.  A \citet{2005MNRAS.362..799M} (M05) or \citet{2003MNRAS.344.1000B} (BC03) single burst model with solar metallicity, a Salpeter IMF and a \citet{2000ApJ...533..682C} extinction law were then fit to these and the IRAC measurements.  The model fluxes were then subtracted from the total flux, and a second model was fit to the remaining flux.  The resulting fits are tabulated in Table \ref{t:fits} and shown in Figure \ref{f:sed}. 	  
	 	
	 The present data do not constrain the peak or shape of the FIR emission, so the total luminosity is uncertain, but the mm and $24\mu$m data imply an infrared luminosity of $0.5-2\times10^{13}L_\sun$, and a corresponding star formation rate (SFR) of $1000-4000M_\sun yr^{-1}$ \citep{2002ApJ...576..159D,2001ApJ...556..562C,1999ApJ...513L..13C}.  The radio flux gives a second estimate of the SFR independent of dust obscuration at $3700\pm700M_\sun yr^{-1}$ assuming a radio spectral slope of -0.8 \citep{1998ARA&A..36..189K}.  Finally, assuming the excess flux in the 3.6$\mu$m band is due to H$\alpha$ originating from the diffuse component, we derive a dust corrected star formation rate of $2900\pm100M_\sun yr^{-1}$ \citep{2000ApJ...533..682C,1998ARA&A..36..189K}, in good agreement with the UV, but lower than the mm, and radio determined values.   This last measurement is largely independent of the models because the H$\alpha$ flux was fixed to the model star formation rate rather than fit as an extra parameter.  
	 	 
	  The bolometric luminosity of this source appears to be dominated by star formation.  The UV and radio morphology suggest a star formation rate density of $15-50M_\sun yr^{-1} kpc^{-2}$, within the range of locally observed starbursts \citep{2005ARA&A..43..677S, 1996ARA&A..34..749S}.  However, without a high resolution map of the mm emission and gas it is not possible to form a clear picture of how the star formation is distributed.  No X-ray flux is detected in a 200ks Chandra exposure placing the X-ray to FIR luminosity ratio in the star formation dominated regime \citep{2005ApJ...632..736A,2003AJ....125..383A}, and the radio to FIR flux ratio falls on the local starburst relation \citep{2001ApJ...554..803Y}.  However, the limit on the X-ray to radio luminosity ratio does not rule out an AGN . The SFR inferred by the diffuse UV and optical emission can explain the FIR emission if the two are spatially related.    Finally, the Ly-$\alpha$ line is narrow and AGN emission lines such as broad C IV or N V are not observed in the optical spectra, so any AGN must either be heavily obscured and/or outside the spectrograph slit.
	  
\section{Implications for galaxy formation}
This object is a likely progenitor for the massive ($>10^{11}$M$_\sun$) old ($>2$Gyr) early type systems seen in large numbers at $z\sim2$ \citep{2007ApJ...669..241M,2006ApJ...638...72K, 2005ApJ...626..680D,2004Natur.430..184C}.  The morphology and spectral properties of the passive galaxies indicate they formed in a single burst at $z>4$ \citep{2008arXiv0801.1184C, 2008ApJ...672..146S,2005ApJ...626..680D}. However, the density of passive $z\sim2$ systems is $\sim10^{-4} Mpc^{-3}$ \citep{2006ApJ...638...72K,2005ApJ...626..680D}, which is too high to be explained by the previous mm source redshift  distribution \citep{2008arXiv0801.1184C}.  

The discovery of this object and other recent studies suggest the fraction of $z>4$ mm sources may be higher than previously thought.   An 850$\mu$m flux limited sample is equivalent to a star formation rate limited sample at $0.5<z<7.5$ and the $z\sim2$ objects must have formed by $z\sim4$ in order to have sufficient time to evolve into passive systems.  With this redshift range and a star formation duration (duty cycle) of 50Myr, a surface density of $\sim200$ objects per square degree is required for these sources to be progenitors of the $z\sim2$ passive galaxy population.  Objects brighter than 4mJy at 850$\mu$m would have a sufficiently high SFR to form $>10^{11}$M$_\sun$ passive systems within 50Myr and the density of such sources is sufficient to form the $z\sim2$ passive galaxies if $\sim30\%$ of them are at $z>4$ \citep{2003MNRAS.344..385B}.  This fraction is well within the range recent studies place at $z>4$ \citep{2007ApJ...671.1531Y, 2006ApJ...647...74W, 2005ApJ...622..772C} .  

\acknowledgements
Support for this work was provided by the Spitzer Science Center which is operated by the Jet Propulsion Laboratory (JPL), California Institute of Technology under NASA contract 1407, NASA through contract 1278386 issued by the JPL and NASA grant HST-GO-09822.  CC thanks the Max-Planck-Gesellschaft and the Humboldt-Stiftung for support through the Max-Planck-Forschungspreis.

\bibliography{ms}

\begin{thebibliography}{38}
\expandafter\ifx\csname natexlab\endcsname\relax\def\natexlab#1{#1}\fi

\bibitem[{{Alexander} {et~al.}(2003){Alexander}, {Bauer}, {Brandt},
  {Hornschemeier}, {Vignali}, {Garmire}, {Schneider}, {Chartas}, \&
  {Gallagher}}]{2003AJ....125..383A}
{Alexander et~al.}, 2003, \aj, 125, 383

\bibitem[{{Alexander} {et~al.}(2005){Alexander}, {Bauer}, {Chapman}, {Smail},
  {Blain}, {Brandt}, \& {Ivison}}]{2005ApJ...632..736A}
{Alexander et~al.}, 2005, \apj, 632, 736

\bibitem[{{Aretxaga} {et~al.}(2007){Aretxaga}, {Hughes}, {Coppin}, {Mortier},
  {Wagg}, {Dunlop}, {Chapin}, {Eales}, {Gazta{\~n}aga}, {Halpern}, {Ivison},
  {van Kampen}, {Scott}, {Serjeant}, {Smail}, {Babbedge}, {Benson}, {Chapman},
  {Clements}, {Dunne}, {Dye}, {Farrah}, {Jarvis}, {Mann}, {Pope}, {Priddey},
  {Rawlings}, {Seigar}, {Silva}, {Simpson}, \& {Vaccari}}]{2007MNRAS.379.1571A}
{Aretxaga et~al.}, 2007, \mnras, 379, 1571

\bibitem[{{Blain} {et~al.}(2002){Blain}, {Smail}, {Ivison}, {Kneib}, \&
  {Frayer}}]{2002PhR...369..111B}
{Blain et~al.} 2002, \physrep, 369, 111

\bibitem[{{Borys} {et~al.}(2003){Borys}, {Chapman}, {Halpern}, \&
  {Scott}}]{2003MNRAS.344..385B}
{Borys et~al.}, 2003, \mnras, 344, 385

\bibitem[{{Bruzual} \& {Charlot}(2003)}]{2003MNRAS.344.1000B}
{Bruzual}, G., \& {Charlot}, S. 2003, \mnras, 344, 1000

\bibitem[{{Calzetti} {et~al.}(2000){Calzetti}, {Armus}, {Bohlin}, {Kinney},
  {Koornneef}, \& {Storchi-Bergmann}}]{2000ApJ...533..682C}
{Calzetti et~al.}, 2000, \apj, 533, 682

\bibitem[{{Capak} {et~al.}(2007){Capak}, {Aussel}, {Ajiki}, {McCracken},
  {Mobasher}, {Scoville}, {Shopbell}, {Taniguchi}, {Thompson}, {Tribiano},
  {Sasaki}, {Blain}, {Brusa}, {Carilli}, {Comastri}, {Carollo}, {Cassata},
  {Colbert}, {Ellis}, {Elvis}, {Giavalisco}, {Green}, {Guzzo}, {Hasinger},
  {Ilbert}, {Impey}, {Jahnke}, {Kartaltepe}, {Kneib}, {Koda}, {Koekemoer},
  {Komiyama}, {Leauthaud}, {Lefevre}, {Lilly}, {Liu}, {Massey}, {Miyazaki},
  {Murayama}, {Nagao}, {Peacock}, {Pickles}, {Porciani}, {Renzini}, {Rhodes},
  {Rich}, {Salvato}, {Sanders}, {Scarlata}, {Schiminovich}, {Schinnerer},
  {Scodeggio}, {Sheth}, {Shioya}, {Tasca}, {Taylor}, {Yan}, \&
  {Zamorani}}]{capak-data}
{Capak et~al.} 2007, \apjs, 172, 99

\bibitem[{{Carilli} {et~al.}(2008){Carilli}, {Lee}, {Capak}, {Schinnerer},
  {Lee}, {McCracken}, {Yun}, {Scoville}, {Smol{\v c}i{\'c}}, {Giavalisco}, \&
  A.}]{carilli-stack}
{Carilli et~al.}, 2008, \apjl

\bibitem[{{Carilli} \& {Yun}(1999)}]{1999ApJ...513L..13C}
{Carilli}, C.~L., \& {Yun}, M.~S. 1999, \apjl, 513, L13

\bibitem[{{Chapman} {et~al.}(2005){Chapman}, {Blain}, {Smail}, \&
  {Ivison}}]{2005ApJ...622..772C}
{Chapman et~al.}, 2005, \apj,
  622, 772

\bibitem[{{Chary} \& {Elbaz}(2001)}]{2001ApJ...556..562C}
{Chary}, R., \& {Elbaz}, D. 2001, \apj, 556, 562

\bibitem[{{Cimatti} {et~al.}(2008){Cimatti}, {Cassata}, {Pozzetti}, {Kurk},
  {Mignoli}, {Renzini}, {Daddi}, {Bolzonella}, {Brusa}, {Rodighiero},
  {Dickinson}, {Franceschini}, {Zamorani}, {Berta}, {Rosati}, \&
  {Halliday}}]{2008arXiv0801.1184C}
{Cimatti et~al.}, 2008, ArXiv e-prints, 801

\bibitem[{{Cimatti} {et~al.}(2004){Cimatti}, {Daddi}, {Renzini}, {Cassata},
  {Vanzella}, {Pozzetti}, {Cristiani}, {Fontana}, {Rodighiero}, {Mignoli}, \&
  {Zamorani}}]{2004Natur.430..184C}
{Cimatti et~al.}, 2004, \nat, 430, 184

\bibitem[{{Daddi} {et~al.}(2005){Daddi}, {Renzini}, {Pirzkal}, {Cimatti},
  {Malhotra}, {Stiavelli}, {Xu}, {Pasquali}, {Rhoads}, {Brusa}, {di Serego
  Alighieri}, {Ferguson}, {Koekemoer}, {Moustakas}, {Panagia}, \&
  {Windhorst}}]{2005ApJ...626..680D}
{Daddi et~al.}, 2005, \apj, 626, 680

\bibitem[{{Dale} \& {Helou}(2002)}]{2002ApJ...576..159D}
{Dale}, D.~A., \& {Helou}, G. 2002, \apj, 576, 159

\bibitem[{{Hughes} {et~al.}(1998){Hughes}, {Serjeant}, {Dunlop},
  {Rowan-Robinson}, {Blain}, {Mann}, {Ivison}, {Peacock}, {Efstathiou}, {Gear},
  {Oliver}, {Lawrence}, {Longair}, {Goldschmidt}, \&
  {Jenness}}]{1998Natur.394..241H}
{Hughes et~al.}, 1998, \nat, 394, 241

\bibitem[{{Kennicutt}(1998)}]{1998ARA&A..36..189K}
{Kennicutt}, Jr., R.~C. 1998, \araa, 36, 189

\bibitem[{{Kong} {et~al.}(2006){Kong}, {Daddi}, {Arimoto}, {Renzini},
  {Broadhurst}, {Cimatti}, {Ikuta}, {Ohta}, {da Costa}, {Olsen}, {Onodera}, \&
  {Tamura}}]{2006ApJ...638...72K}
{Kong et~al.}, 2006, \apj, 638, 72

\bibitem[{{Lee} {et~al.}(2008){Lee}, {Giavalisco}, {Capak}, {Scoville},
  {Mobasher}, {Taniguchi}, \& {Sasaki}}]{soo-lbg}
{Lee et~al.}, 2008, \apj

\bibitem[{{Maraston}(2005)}]{2005MNRAS.362..799M}
{Maraston}, C. 2005, \mnras, 362, 799

\bibitem[{{Marinoni} {et~al.}(2001){Marinoni}, {Davis}, {Coil}, \&
  {Finkbeiner}}]{2001astro.ph..9164M}
{Marinoni et~al.}, 2001, ArXiv
  Astrophysics e-prints

\bibitem[{{McGrath} {et~al.}(2007){McGrath}, {Stockton}, \&
  {Canalizo}}]{2007ApJ...669..241M}
{McGrath}, E.~J., {Stockton}, A., \& {Canalizo}, G. 2007, \apj, 669, 241

\bibitem[{{Pettini} {et~al.}(2000){Pettini}, {Steidel}, {Adelberger},
  {Dickinson}, \& {Giavalisco}}]{2000ApJ...528...96P}
{Pettini et~al.}, 2000, \apj, 528, 96

\bibitem[{{Pope} {et~al.}(2005){Pope}, {Borys}, {Scott}, {Conselice},
  {Dickinson}, \& {Mobasher}}]{2005MNRAS.358..149P}
{Pope et~al.}, 2005, \mnras, 358, 149

\bibitem[{{Salpeter}(1955)}]{1955ApJ...121..161S}
{Salpeter}, E.~E. 1955, \apj, 121, 161

\bibitem[{{Sanders} \& {Mirabel}(1996)}]{1996ARA&A..34..749S}
{Sanders}, D.~B., \& {Mirabel}, I.~F. 1996, \araa, 34, 749

\bibitem[{{Sanders} {et~al.}(2007){Sanders}, {Salvato}, {Aussel}, {Ilbert},
  {Scoville}, {Surace}, {Frayer}, {Sheth}, {Helou}, {Brooke}, {Bhattacharya},
  {Yan}, {Kartaltepe}, {Barnes}, {Blain}, {Calzetti}, {Capak}, {Carilli},
  {Carollo}, {Comastri}, {Daddi}, {Ellis}, {Elvis}, {Fall}, {Franceschini},
  {Giavalisco}, {Hasinger}, {Impey}, {Koekemoer}, {Le F{\`e}vre}, {Lilly},
  {Liu}, {McCracken}, {Mobasher}, {Renzini}, {Rich}, {Schinnerer}, {Shopbell},
  {Taniguchi}, {Thompson}, {Urry}, \& {Williams}}]{sanders-spitzer}
{Sanders et~al.}, 2007, \apjs, 172, 86

\bibitem[{{Schinnerer} {et~al.}(2007){Schinnerer}, {Smol{\v c}i{\'c}},
  {Carilli}, {Bondi}, {Ciliegi}, {Jahnke}, {Scoville}, {Aussel}, {Bertoldi},
  {Blain}, {Impey}, {Koekemoer}, {Le Fevre}, \& {Urry}}]{eva-vla}
{Schinnerer et~al.} 2007, \apjs, 172, 46

\bibitem[{{Scott} {et~al.}(2008){Scott}, {Austermann}, {Perera}, {Wilson},
  {Aretxaga}, {Bock}, {Hughes}, {Kang}, {Kim}, {Mauskopf}, {Sanders},
  {Scoville}, \& {Yun}}]{2008arXiv0801.2779S}
{Scott et~al.}, 2008, \mnras, 385, 2225

\bibitem[{{Scoville} {et~al.}(2007){Scoville}, {Abraham}, {Aussel}, {Barnes},
  {Benson}, {Blain}, {Calzetti}, {Comastri}, {Capak}, {Carilli}, {Carlstrom},
  {Carollo}, {Colbert}, {Daddi}, {Ellis}, {Elvis}, {Ewald}, {Fall},
  {Franceschini}, {Giavalisco}, {Green}, {Griffiths}, {Guzzo}, {Hasinger},
  {Impey}, {Kneib}, {Koda}, {Koekemoer}, {Lefevre}, {Lilly}, {Liu},
  {McCracken}, {Massey}, {Mellier}, {Miyazaki}, {Mobasher}, {Mould}, {Norman},
  {Refregier}, {Renzini}, {Rhodes}, {Rich}, {Sanders}, {Schiminovich},
  {Schinnerer}, {Scodeggio}, {Sheth}, {Shopbell}, {Taniguchi}, {Tyson}, {Urry},
  {Van Waerbeke}, {Vettolani}, {White}, \& {Yan}}]{scoville-hst}
{Scoville et~al.}, 2007, \apjs, 172, 38

\bibitem[{{Shapley} {et~al.}(2003){Shapley}, {Steidel}, {Pettini}, \&
  {Adelberger}}]{2003ApJ...588...65S}
{Shapley et~al.}, 2003,
  \apj, 588, 65

\bibitem[{{Solomon} \& {Vanden Bout}(2005)}]{2005ARA&A..43..677S}
{Solomon}, P.~M., \& {Vanden Bout}, P.~A. 2005, \araa, 43, 677

\bibitem[{{Stockton} {et~al.}(2008){Stockton}, {McGrath}, {Canalizo}, {Iye}, \&
  {Maihara}}]{2008ApJ...672..146S}
{Stockton et~al.},
  2008, \apj, 672, 146

\bibitem[{{Wang} {et~al.}(2006){Wang}, {Cowie}, \&
  {Barger}}]{2006ApJ...647...74W}
{Wang}, W.-H., {Cowie}, L.~L., \& {Barger}, A.~J. 2006, \apj, 647, 74

\bibitem[{{Wang} {et~al.}(2007){Wang}, {Cowie}, {van Saders}, {Barger}, \&
  {Williams}}]{2007ApJ...670L..89W}
{Wang et~al.}, 2007, \apjl, 670, L89

\bibitem[{{Wilson} {et~al.}(2008){Wilson}, {Austermann}, {Perera}, {Scott},
  {Ade}, {Bock}, {Glenn}, {Golwala}, {Kim}, {Kang}, {Lydon}, {Mauskopf},
  {Predmore}, {Roberts}, {Souccar}, \& {Yun}}]{2008arXiv0801.2783W}
{Wilson et~al.}, 2008, \mnras, 386, 807

\bibitem[{{Younger} {et~al.}(2007){Younger}, {Fazio}, {Huang}, {Yun}, {Wilson},
  {Ashby}, {Gurwell}, {Lai}, {Peck}, {Petitpas}, {Wilner}, {Iono}, {Kohno},
  {Kawabe}, {Hughes}, {Aretxaga}, {Webb}, {Mart{\'{\i}}nez-Sansigre}, {Kim},
  {Scott}, {Austermann}, {Perera}, {Lowenthal}, {Schinnerer}, \& {Smol{\v
  c}i{\'c}}}]{2007ApJ...671.1531Y}
{Younger et~al.}, 2007, \apj, 671, 1531

\bibitem[{{Yun} {et~al.}(2001){Yun}, {Reddy}, \&
  {Condon}}]{2001ApJ...554..803Y}
{Yun}, M.~S., {Reddy}, N.~A., \& {Condon}, J.~J. 2001, \apj, 554, 803

\end{thebibliography}

\end{document}